\documentclass[fleqn,pra,twocolumn]{revtex4}

\usepackage{amsmath}
\usepackage{dcolumn}
\usepackage{multirow}
\usepackage{booktabs}
\usepackage{verbatim}

\newcommand{\eq}{Eq.~}
\newcommand{\se}{SE}
\newcommand\Tstrut{\rule{0pt}{2.5ex}}         % = `top' strut
\newcommand\Bstrut{\rule[-1.3ex]{0pt}{0pt}}   % = `bottom' strut
\newcommand{\footstar}[1]{$^*$ \footnotetext{$^*$#1}}

\begin{document}

\title{One-loop self-energy using a numerical Green function}

\author{Hugo D. Nogueira$^{1,}$\footstar{Present address: Laboratoire de Chimie Th\'eorique, Sorbonne Universit\'e, UMR 7616 CNRS, F-75005 Paris, France}, Maen Salman$^{1,*}$ and Jean-Philippe Karr$^{1,2}$}
\affiliation{$^1$Laboratoire Kastler Brossel, Sorbonne Universit\'e, CNRS, ENS-Universit\'e PSL, Coll\`ege de France, 4 place Jussieu, F-75005 Paris, France}
\affiliation{$^2$Universit\'e d'Evry-Val d'Essonne, Universit\'e Paris-Saclay, Boulevard Fran\c cois Mitterrand, F-91000 Evry, France}

\begin{abstract}

We calculate the one-loop self-energy in hydrogenlike atoms using a numerical Green function obtained by solving the radial Dirac equation in an exponential basis set. The self-energy correction in the ground state of hydrogenlike uranium is obtained with about $10^{-5}$ relative uncertainty in the Feynman gauge. Using a convergence acceleration scheme, we extend our calculations to the region of low nuclear charges. Our results allow calculating the self-energy correction for the hydrogen atom with $10^{-4}$ relative uncertainty. Calculations in the Coulomb gauge are also presented, improving the precision to $10^{-5}$. Present limitations and possible improvements of our method are discussed. 

\end{abstract}

\maketitle

\section{Introduction} \label{sec-intro}

The bound-electron self-energy, in particular its leading one-loop component, is the largest quantum electrodynamics (QED) correction to the energy levels of atoms and molecules. Since the very first approximate evaluation of the one-loop self-energy (SE) in the hydrogen atom allowed to understand the experimentally observed Lamb shift~\cite{Bethe:1947,Lamb:1947}, calculations in hydrogen-like systems have been continuously improved using two different approaches. In the first one, calculations are performed in a nonrelativistic framework and the nuclear binding field is treated perturbatively, leading to an expansion in powers of $Z\alpha$, where $Z$ is the nuclear charge and $\alpha$ the fine-structure constant (see~\cite{Eides:2007,Yerokhin:2019,Mohr:2025} for reviews). The second one is a direct numerical evaluation of the energy shift as obtained from relativistic QED theory~\cite{Mohr:1998,Indelicato:2019}, amounting to an exact all-order treatment of the electron-nucleus interaction. The nonrelativistic approach is well suited for low-$Z$ systems, whereas the relativistic one has been mostly used in heavy atoms. Nevertheless, an extraordinary precision level of $10^{-10}$ was reached in all-order calculations for the hydrogen atom ($Z=1$)~\cite{Jentschura:1999,Jentschura:2001}, surpassing that of the nonrelativistic $Z\alpha$-expansion, which has been pushed to the orders $mc^2\alpha (Z\alpha)^6$~\cite{Jentschura:2005} and 
$mc^2\alpha (Z\alpha)^7\ln(\alpha)$~\cite{Karshenboim:1997}.

This suggests that the numerical approach might be a promising way to improve \se{}  calculations in light non-hydrogenic systems as well. In particular, the molecular hydrogen ions (H$_2^+$, HD$^+$, $\ldots$) are studied in laser spectroscopy experiments at rapidly increasing levels of accuracy. A couple of measurements~\cite{Patra:2020,Alighanbari:2023} are now more precise than the theoretical predictions~\cite{Korobov:2017,Korobov:2021}. The largest source of theoretical uncertainty is the \se{} correction, which has been calculated using the nonrelativistic expansion method up to the same order as in the hydrogenic case~\cite{Korobov:2017,Korobov:2021,Korobov:2014a,Korobov:2014b}, reaching close to $10^{-6}$ precision.

Coming back to the hydrogen atom case, it is important to note that the high-precision results of Refs.~\cite{Jentschura:1999,Jentschura:2001}, which are based on an approach developed by Mohr~\cite{Mohr:1974a,Mohr:1974b}, require a large amount of computing power. The all-order approach relies on a partial-wave decomposition of the Dirac-Coulomb propagator, and the slow convergence of this expansion is one of the main numerical challenges for highly accurate calculations. In~\cite{Jentschura:1999,Jentschura:2001}, this was resolved by using convergence acceleration techniques, which still required the calculation of partial-wave contributions as high as a few millions. Because of this, extension of this approach to the more complex molecular case appears as a daunting task.

An alternative method, which is now the most widely used one, is the so-called potential-expansion method, introduced in~\cite{Snyderman:1991,Blundell:1991} and further developed into a highly efficient numerical procedure in~\cite{Yerokhin:1999}. In this approach, the first two terms of the potential expansion are subtracted from the \se{} operator, leading to a separation into zero-, one-, and many-potential terms. The zero- and one-potential terms are renormalized and evaluated in momentum space, while the many-potential term, the only one that requires a partial-wave expansion, is the numerically most difficult term. However, subtraction techniques have been devised to accelerate the convergence of the partial-wave expansion~\cite{Yerokhin:2005,Sapirstein:2023}. An additional difficulty comes from large numerical cancellations occurring for small values of $Z$, due to the existence of spurious terms in the individual contributions in the Feynman gauge in which the theory was initially formulated. It was shown that use of the Coulomb gauge~\cite{Adkins:1983,Adkins:1986} avoids this problem and yields significant accuracy improvements~\cite{Hedendahl:2012,Yerokhin:2025}. For example, in~\cite{Yerokhin:2025} the hydrogen atom ground state correction was evaluated with precision better than $10^{-7}$. These advances open a realistic route towards extension of \se{} calculations to light non-hydrogenic systems.

Calculations in molecular hydrogen ions, where the electron is bound by a two-center potential, are more complex for two main reasons. One is the loss of spherical symmetry, replaced by an axial one, and the other is that the Dirac-Coulomb Green function is not known in analytical form. This means that the \se{} calculations have to be done with a numerical Green function, expressed as a sum over the Dirac eigenvectors and eigenvalues obtained by a finite basis set method~\cite{Yerokhin:2020}. By comparison, all the above-mentioned high-precision results made use of the expression of the hydrogenic Dirac Green function in terms of Whittaker functions~\cite{Mohr:1998,Mohr:1974a,Mohr:1974b}. \se{} calculations with numerical Green functions were performed in a few works, using either the $B$-spline method~\cite{Blundell:1991,Cheng:1993,Shabaev:2004} or a Gaussian basis set~\cite{Ferenc:2025}. Notably, the $B$-spline method was also applied to heavy diatomic molecular ions~\cite{Artemyev:2015,Artemyev:2022}. However, these works only considered heavy systems, and the accuracy of the total \se{} correction did not exceed the five-digit level.

The goal of the present work is to demonstrate the feasibility of high-precision \se{} calculations in light of one-electron systems using a numerical Green function. As a benchmark, we study the ground $1s_{1/2}$ state of hydrogenic atoms, focusing on the many-potential term, which is the most challenging numerically. In the perspective of future extension to molecular hydrogen ions, we use a basis set of exponential functions, constructed in a similar way to that used by us in Ref.~\cite{Nogueira:2023} to solve the Dirac equation for molecular ions. In that work, this basis set was shown to yield highly accurate bound-state energies, and the accuracy of the numerical Green function was also tested by evaluating sum rules.

The paper is organized as follows. The theoretical expressions of the self-energy correction, as well as the calculation of the numerical Green function and the convergence acceleration scheme are reviewed in Sec.~\ref{sec-se-eq}. The approach we use to solve the radial Dirac equation in hydrogenic systems is described in Sec.~\ref{sec-dirac-eq}. We give some details on our numerical approach for the SE calculation in Sec.~\ref{sec-num}, and present our results in Sec.~\ref{sec-results}.

\section{Mathematical formulation} \label{sec-se-eq}

\subsection{Expressions of the self-energy correction}

Using relativistic units ($\hbar=c=m=1$), the energy shift for an electron in a bound state $\psi_a$ caused by the \se{} is given by \cite{Yerokhin:1999}
\begin{equation}\label{eq_se1}
\begin{split}
\Delta E_{\rm SE}=&2i\alpha\! \int_{C_{\rm F}}\!dz\!
\int \!d^3\textbf{x}_2 \!\int \!d^3\textbf{x}_1 
\psi_a^{\dagger}(\textbf{x}_1)
\alpha_{\mu} 
G(\textbf{x}_1,\textbf{x}_2,E_a\!-\!z) \\
& \alpha_{\nu}
\psi_a(\textbf{x}_2)
D^{\mu\nu}(\textbf{x}_{12},z) 
- \delta m\int d^3\textbf{x} \,
\overline{\psi}_a(\textbf{x}) 
\psi_a(\textbf{x})\,,
\end{split}
\end{equation}
where $C_F$ is the standard Feynman integration contour in the complex plane, $\alpha_{\mu}=(\rm I, \boldsymbol{\alpha})$ are the Dirac matrices, $G(\textbf{x}_1,\textbf{x}_2,z)$ is the Dirac Green function, $\delta m$ the mass counterterm, the bar over the wavefunction denotes the Dirac adjoint ($\overline{\psi}_a(\textbf{x})=\psi_a^{\dagger}(\textbf{x})\gamma_0$), and $D^{\mu\nu}(\textbf{x}_{12},z)$ the photon propagator in the chosen gauge. The bound solution of the hydrogenic Dirac equation may be written as
\begin{equation} \label{eq_psi_a}
\psi_a(\mathbf{x}) = \begin{pmatrix}
    \phi_a^1 (x) \, \eta_{\kappa}^{\mu} (\mathbf{\hat{r}})\\
    i\phi_a^2 (x) \, \eta_{-\kappa}^{\mu} (\mathbf{\hat{r}})
\end{pmatrix},
\end{equation}
where $\eta_{\kappa}^{\mu} (\mathbf{\hat{r}})$ is a two-component spin-angular spinor (see, e.g.,~\cite{Mohr:1974a}).

Expression (\ref{eq_se1}) contains divergences, and therefore cannot be directly evaluated. One way of isolating the ultraviolet divergences of \eq(\ref{eq_se1}) is to expand the electron Green function in powers of the potential $V$~\cite{Snyderman:1991,Yerokhin:1999}, using the identity
\begin{equation}\label{eq_pov1}
G = G_0 + G_0VG_0 + G_0VG_0VG_0 + \ldots \,,
\end{equation}
Here, $G=1/(z-H)$, with $H$ being the Dirac-Coulomb Hamiltonian, and $G_0=1/(z-H_0)$, where $H_0$ is the free Dirac Hamiltonian. As a result, the \se{} correction is  divided into the so-called zero-, one-, and many-potential terms. All the divergences are contained in the zero- and one-potential terms, which require renormalization. Their final renormalized expressions are written in the form
\begin{equation}\label{eq_pov9}
\Delta E_{\rm SE}^0 = \int \frac{\,d^3\textbf{p}}{(2\pi)^3}
\overline{\psi}_a(\textbf{p})
\Sigma^0_R({\rm p})
\psi_a(\textbf{p})\,,
\end{equation}
and 
\begin{equation}\label{eq_pov13}
\begin{split}
\Delta E_{\rm SE}^1 = \int \frac{\,d^3\textbf{p}}{(2\pi)^3}
\int \frac{\,d^3\textbf{p}^{\prime}}{(2\pi)^3}
&\overline{\psi}_a(\textbf{p}^{\prime})
\Gamma^0_R({\rm p}^{\prime},{\rm p}) \\
&V(|\textbf{p}^{\prime}-\textbf{p}|)
\psi_a(\textbf{p})\,,
\end{split}
\end{equation}
where $\psi_a(\textbf{p})$ is the momentum space Dirac wavefunction, and the expressions of the operators $\Sigma^0_R({\rm p})$ and $\Gamma^0_R({\rm p}^{\prime},{\rm p})$ in the Feynman and Coulomb gauges are given in Refs.~\cite{Yerokhin:1999} and \cite{Yerokhin:2025}, respectively.

The many-potential term is finite and does not require renormalization. After integrating analytically over angles, it is given in the Feynman gauge by~\cite{Yerokhin:1999}
\begin{equation}\label{eq_mpp1}
\begin{split}
&\Delta E_{\rm SE,F}^{2+} = \frac{i\alpha}{2\pi} \frac{1}{2j_a+1} 
\sum_{\kappa=-\infty}^{\infty} 
\int_0^{\infty}\,dx_1 x_1^2\int_0^{\infty}\,dx_2 x_2^2 \\
&\int_{C_L+C_H}\,dz \times \sum_J (2J+1)
\left\{ \vphantom{\sum_{L=J-1}^{J+1}}  [C_J(\kappa,\kappa_a)]^2 g_J(z) A_{\kappa}^{2+} \right. \\
& \hspace{4cm} \left.-\sum_{L=J-1}^{J+1} g_L(z) B_{\kappa,J,L}^{2+}  \right\}\,,
\end{split}
\end{equation}
where
\begin{subequations} \label{eq_AB}
\begin{equation}
A_{\kappa}^{2+} = \sum_{i,j=1}^2 
\phi_a^i(x_1) 
[G_{\kappa}^{2+}]_{i,j} (x_1,x_2,E_a-z)
\phi_a^j(x_2)\,,
\end{equation}
\begin{equation}
B_{\kappa,J,L}^{2+} = \sum_{i,j=1}^2 
\phi_a^{\overline{i}}(x_1) 
[G_{\kappa}^{2+}]_{i,j} (x_1,x_2,E_a-z)
\phi_a^{\overline{j}}(x_2)
S_{\kappa,i,j}^{J,L}\,,
\end{equation}
\end{subequations}
In the above expression, $\overline{i}=3-i$, and $S_{\kappa,i,j}^{J,L}$ are given by
\begin{equation}
\begin{split}
& S_{\kappa,1,1}^{J,L} = S_{J,L}(-\kappa_a,\kappa) S_{J,L}(-\kappa_a,\kappa) \,,\\
& S_{\kappa,1,2}^{J,L} = -S_{J,L}(-\kappa_a,\kappa) S_{J,L}(\kappa_a,-\kappa) \,,\\
& S_{\kappa,2,1}^{J,L} = -S_{J,L}(\kappa_a,-\kappa) S_{J,L}(-\kappa_a,\kappa) \,,\\
& S_{\kappa,2,2}^{J,L} = S_{J,L}(\kappa_a,-\kappa) S_{J,L}(\kappa_a,-\kappa) \,.\\
\end{split}
\end{equation}
Finally, the $g_l(z)$, $C_J(\kappa_a,\kappa_b)$ and $S_{J,L}(\kappa_a,\kappa_b)$ coefficients are given in the Appendix C of~\cite{Yerokhin:1999}.

In \eq (\ref{eq_AB}), $G_{\kappa}^{2+}$ is the radial part of the term of the $\kappa$ expansion of $G^{2+}$, where $G^{2+}$ is the part of the electron propagator containing two or more Coulomb interactions (see Appendix~D in~\cite{Yerokhin:1999} for details). With a potential expansion similar to \eq(\ref{eq_pov1}), $G_{\kappa}^{2+}$ can be obtained by the subtraction~\cite{Yerokhin:1999}
\begin{equation} \label{eq_g2p}
\begin{split}
G_{\kappa}^{2+}(x_1,x_2,z) =& \; G_{\kappa} (x_1,x_2,z) \\
&- G_{\kappa}^0 (x_1,x_2,z) - G_{\kappa}^1 (x_1,x_2,z)\,,
\end{split}
\end{equation}
where $G_{\kappa}$ and $G_{\kappa}^0$ are the radial Green functions for a bound and free electron, respectively, and $G_{\kappa}^1$ is given by
\begin{equation} \label{eq_g1}
G_{\kappa}^{1} (x_1,x_2,z) = \int_0^{\infty} dx \, x^2 G_{\kappa}^{0} (x_1,x,z) \, V(x) \, G_{\kappa}^{0} (x,x_2,z).
\end{equation}

In the Coulomb gauge, the expression of the many-potential term is~\cite{Yerokhin:2025}
\begin{equation}\label{eq_mpp1_coulomb}
\begin{split}
&\Delta E_{\rm SE,C}^{2+} = \frac{i\alpha}{2\pi} \frac{2}{2j_a+1} 
\sum_{\kappa=-\infty}^{\infty} 
\int_0^{\infty}dx_1 \, x_1^2\int_0^{\infty} dx_2 \, x_2^2 \\
&\int_{C_L+C_H}\,dz \times \sum_J 
\left[ \vphantom{\sum_{L=J-1}^{J+1}} (2J+1) [C_J(\kappa,\kappa_a)]^2 g_J(0) A_{\kappa}^{2+} \right.\\
& -\sum_{L=J-1}^{J+1} a_{JL} g_L(z) D_{\kappa,J,L,L}^{2+} \\
& +\sqrt{J(J+1)} g_{J}^{ret,1}(z) D_{\kappa,J,J+1,J-1}^{2+} \\
& \left. +\sqrt{J(J+1)} g_{J}^{ret,2}(z) D_{\kappa,J,J-1,J+1}^{2+} \right]\,.
\end{split}
\end{equation}
The expressions of $a_{JL}$, $g_J^{ret,1}(z)$, $g_J^{ret,2}(z)$ and $D^{2+}_{\kappa,J,L,L'}$ can be found in~\cite{Yerokhin:2025}. In this work, we use the integration contour described in~\cite{Yerokhin:2005}. The low-energy part $C_L$ extends from $\varepsilon_0 - i0$ to $-i0$ on the lower side of the branch cut of the photon propagator, and from $+i0$ to $\varepsilon_0 + i0$ on the upper side, with $\varepsilon_0 = Z\alpha E_a$. Since we only perform calculations in the ground state, it is not necessary to bend the contour in the complex plane to avoid the poles of the electron propagator. The high-energy part $C_H$ is $]\varepsilon_0 - i\infty,\varepsilon_0 - i0]+[\varepsilon_0 + i0,\varepsilon_0 + i\infty[$.

\subsection{Numerical Green functions} \label{sec-Green}

The most important ingredient for calculation of the many-potential term is the radial Green function, more precisely the part that involves two or more interactions with the potential, $G_{\kappa}^{2+}(x_1,x_2,z)$ (see Eqs.~(\ref{eq_AB}) and (\ref{eq_g2p})). In this work, we use a numerical Green function given by
\begin{equation} \label{gf1}
[G_{\kappa}]_{i,j} (x_1,x_2,z) = \sum_n \frac{\phi_{\kappa,n}^i (x_1) \phi_{\kappa,n}^j(x_2)}{z - E_{\kappa,n}} \,,
\end{equation}
where $n$ runs over all solutions of the radial Dirac equation, which are obtained by the numerical method described in Sec.~\ref{sec-dirac-eq}.

Using \eq(\ref{eq_g2p}) in \eq(\ref{eq_mpp1}), a given $\kappa$ contribution of the many-potential term is evaluated as
\begin{equation} \label{eq_Emany2p}
\Delta E_{\kappa}^{2+} = \Delta E_{\kappa} - \Delta E_{\kappa}^{0} - \Delta E_{\kappa}^{1} \,,
\end{equation}
where $\Delta E$, $\Delta E^{0}$, and $\Delta E^{1}$ are given by \eq(\ref{eq_mpp1}) or (\ref{eq_mpp1_coulomb}), with $G_{\kappa}^{2+}$ in \eq(\ref{eq_AB}) respectively substituted with the full Green function $G_{\kappa}$ of \eq(\ref{gf1}), with the free Green function $G_{\kappa}^{0}$, and with $G_{\kappa}^{1}$ [\eq(\ref{eq_g1})], which we refer to as the one-potential Green function.

Although the free Green function $G_{\kappa}^{0}$ is known in analytical form~\cite{Mohr:1974a}, it is advantageous to calculate it numerically by solving the free Dirac equation in the same basis set as that used for the hydrogenic problem - the only difference being that the nuclear charge $Z$ is set to zero - and using the resulting eigenvalues and eigenvectors in \eq(\ref{gf1}). Indeed, this favors cancellation of numerical errors in the subtraction in \eq(\ref{eq_Emany2p})~\cite{Ferenc:2025}, see Sec.~\ref{sec-results}.

There are different possible ways to calculate $G_{\kappa}^{1}$~\cite{Nogueira:2024}. It may be obtained from a numerical derivative of $G_{\kappa}$ over $Z$~\cite{Yerokhin:1999}. In order to avoid precision loss from this procedure, we instead calculate it by analytical integration of \eq(\ref{eq_g1})~\cite{Nogueira:2024,Ferenc:2025}. The calculation of $G_{\kappa}^{1}$ is further detailed in Sec.~\ref{sec-g1}.

\subsection{Convergence acceleration} \label{sec_accel}

In order to accelerate the convergence of the partial-wave expansion in \eq(\ref{eq_mpp1}) or \eq(\ref{eq_mpp1_coulomb}), we use the acceleration scheme demonstrated in~\cite{Sapirstein:2023}. The general idea is to subtract and add back an approximation of the many-potential Green function, $G^{2+}$, that is both calculable in a closed form (without partial-wave expansion) and captures the slowest-converging part of the $\kappa$-expansion. We subtract from each $\kappa$ component $\Delta E_{\kappa}^{2+}$ of \eq(\ref{eq_mpp1}) or \eq(\ref{eq_mpp1_coulomb}) the quantity $\Delta E_{\rm substr,\kappa}^{2+}$, which has the same expression as $\Delta E_{\kappa}^{2+}$ except that $G_{\kappa}^{2+}$ is substituted with the following approximate form that corresponds to the two-potential term with the potentials commuted outside the integral~\cite{Sapirstein:2023,Yerokhin:2025}:
\begin{equation}
    G_{a,\kappa}^{2+}(x_1,x_2,z) = \frac{1}{2}V(x_1)\frac{\partial^2}{\partial z^2}G_{\kappa}^{0}(x_1,x_2,z)V(x_2)\,,
\end{equation}
which can be obtained analytically. One then has to add back the quantity
\begin{equation}
\Delta E_{\rm substr}^{2+} = \sum_{\kappa=-\infty}^{\infty} \Delta E_{\rm substr,\kappa}^{2+}\,,
\end{equation}
which can be expressed in momentum space as~\cite{Yerokhin:2025}
\begin{equation} \label{eq_deltaES}
    \Delta E_{\rm substr}^{2+}=\frac{1}{2}\int\frac{d^3\mathbf{p}}{(2\pi)^3}\overline{\psi_V}(\mathbf{p}) \left. \frac{\partial^2\Sigma_R^{(0)}(\varepsilon,\mathbf{p})}{\partial \varepsilon^2} \right|_{\varepsilon = E_a} \!\psi_V(\mathbf{p})\,,
\end{equation}
where
\begin{equation}
    \psi_V(\mathbf{p}) = \int d^3\mathbf{x} \, e^{-i\mathbf{p}\cdot\mathbf{x}}V(\mathbf{x})\psi_a(\mathbf{x})\,.
\end{equation}

\section{Dirac equation and basis sets} \label{sec-dirac-eq}

We present here our numerical method to solve the Dirac equation in hydrogenic systems, which allows us to construct the Green functions discussed in Sec.~\ref{sec-Green}. In this Section we switch to atomic units ($\hbar = m = e = 1$). The Dirac equation is
\begin{subequations} \label{eq_dirac}
\begin{flalign}
&H \psi_n = E_n \psi_n, \\
&H = (\beta\!-\!I_4) c^2 + c \boldsymbol{\alpha}\mathbf{p} + V = 
    \begin{pmatrix}
      V & c \boldsymbol{\sigma}\mathbf{p} \\
      c \boldsymbol{\sigma}\mathbf{p}  & V \!-\! 2 c^2
    \end{pmatrix},&
\end{flalign}
\end{subequations}
where $H$ is the Dirac Hamiltonian, $\psi_n, E_n$ are the Dirac wavefunctions and energies. In \eq(\ref{eq_dirac}b), $\boldsymbol{\alpha}$ and $\beta$ are the 4$\times$4 Dirac matrices, $V=-Z/x$ is the Coulomb potential, $\mathbf{p}$ is the momentum operator, and $\boldsymbol{\sigma}$ are the Pauli matrices.

We write the solution of the Dirac equation similarly to \eq(\ref{eq_psi_a}):
\begin{equation} \label{eq_psi_n}
\psi_{\kappa,n} (\mathbf{x}) = \begin{pmatrix}
    \phi_{\kappa,n}^1 (x) \, \eta_{\kappa}^{\mu} (\mathbf{\hat{r}})\\
    i\phi_{\kappa,n}^2 (x) \, \eta_{-\kappa}^{\mu} (\mathbf{\hat{r}})
\end{pmatrix}.
\end{equation}
Injecting this expression into \eq(\ref{eq_dirac}a) yields the well-known radial Dirac equation
\begin{equation} \label{eq_Dirac_rad}
\begin{pmatrix}
V & c\left( -\frac{d}{dx} \!+\! \frac{\kappa}{x} \right) \\
c\left( \frac{d}{dx} \!+\! \frac{\kappa}{x} \right)  & V - 2 c^2 
\end{pmatrix}
\begin{pmatrix}
P_{\kappa,n}(x) \\
Q_{\kappa,n}(x)
\end{pmatrix}
= E_{\kappa,n}
\begin{pmatrix}
P_{\kappa,n}(x) \\
Q_{\kappa,n}(x)
\end{pmatrix}
\end{equation}
with $P_{\kappa,n}(x) = x \phi_{\kappa,n}^1(x)$, $Q_{\kappa,n}(x) = x \phi_{\kappa,n}^2(x)$.

The radial functions $P_{\kappa,n}$ and $Q_{\kappa,n}$ are expanded in a finite basis set. In a first approach, which we denote by NKB~\cite{Nogueira:2023}, no kinetic balance conditions are imposed between the basis functions $\pi_{\mu}^L$ and $\pi_{\mu}^S$, which respectively describe the large ($P_{\kappa,n}$) and small ($Q_{\kappa,n}$) components. The expansion is then written as
\begin{equation}
P_{\kappa,n}(x) = \sum_{\mu = 1}^{N}  A_{\mu} \pi_{\mu}^L(x) \, , \;\; Q_{\kappa,n}(x) = \sum_{\mu = 1}^{N} B_{\mu} \pi_{\mu}^S(x) \,.
\label{eq-basis-exp}
\end{equation}
A second approach is the widely used restricted kinetic balance (RKB) scheme~\cite{Stanton:1984}, where the basis functions describing the small component are given by the relationship
\begin{equation} \label{eq_basis_rkb}
\pi_{\mu}^S = \frac{1}{2c} \left[ \frac{d}{dx} + \frac{\kappa}{x} \right] \pi_{\mu}^L\,.
\end{equation}
Finally, we performed calculations using the dual kinetic balance (DKB) scheme, which was shown to improve the convergence of \se{} calculations~\cite{Shabaev:2004}:
\begin{equation} \label{eq_basis_dkb}
\begin{split}
\begin{pmatrix}
P_{\kappa,n} \\
Q_{\kappa,n}
\end{pmatrix} = \sum_{\mu=1}^{N} &\left[
A_{\mu}
\begin{pmatrix}
\pi_{\mu}^L \\
\frac{1}{2c} \left[ \frac{d}{dx} \!+\! \frac{\kappa}{x} \right] \pi_{\mu}^L
\end{pmatrix} \right. \\
&+ \left.
B_{\mu}
\begin{pmatrix}
\frac{1}{2c} \left[ \frac{d}{dx} \!-\! \frac{\kappa}{x} \right] \pi_{\mu}^S \\
\pi_{\mu}^S
\end{pmatrix}
\right] .
\end{split}
\end{equation}
The matrix form of the Dirac equation for the three above-described kinetic balance schemes is given in Appendix~\ref{app_Dirac}.

Having in mind a future extension of our numerical method to molecular hydrogen ions, we use an exponential basis set, similar to that used to solve the two-center Dirac~\cite{Nogueira:2023} equation. The basis functions $\pi_{\mu}^L$, as well as $\pi_{\mu}^S$ (except in the RKB scheme where $\pi_{\mu}^S$ is given by \eq(\ref{eq_basis_rkb})), are
\begin{equation}\label{eq_nkbhlike1}
\pi_\mu(x) = x^{\ell+1} e^{-\alpha_{\mu} x}\,,
\end{equation}
where $\ell$ is the orbital momentum quantum number, and the exponents (nonlinear parameters) $\alpha_{\mu}$ are generated pseudorandomly in several intervals. Similarly to Ref.~\cite{Nogueira:2023}, high exponents are included to better represent the singular behavior of the Dirac wavefunction at the origin. For example values of the interval bounds, the reader may consult Fig.~2 of~\cite{Korobov:2012}, where a similar basis set was used for the Schr\"odinger equation in the context of Bethe logarithm calculations.

The exponential basis set has the following noteworthy property when applied to the hydrogenlike atom problem. The RKB prescription~(\ref{eq_basis_rkb}), applied to a large-component basis function of the type~(\ref{eq_nkbhlike1}), yields
\begin{equation}
\pi_{\mu}^S = \frac{1}{2c} \left[-\alpha_{\mu} + \frac{\ell+1+\kappa}{x} \right] \pi_{\mu}^L\,.
\end{equation}
For $\kappa<0$, the second term in the square brackets vanishes, so that $\pi_{\mu}^S \propto \pi_{\mu}^L$. This shows that the NKB and RKB schemes are strictly equivalent for $\kappa<0$ states. However, this is not the case for $\kappa>0$ states.

\section{Numerical implementation} \label{sec-num}

\subsection{Numerical details}

The above-described method was implemented in Fortran 90. The resolution of the Dirac equation and three-dimensional integration involved in the \se{} calculation (see Eqs.~(\ref{eq_mpp1}) and (\ref{eq_mpp1_coulomb})) are carried out with high-precision arithmetic using the package MPFUN2020~\cite{Bailey:2024}. The diagonalization of the Dirac Hamiltonian and the \se{} integration were parallelized using the OpenMP programming interface.

The integrals over $x_1$ and $x_2$ in Eqs.~(\ref{eq_mpp1}) and (\ref{eq_mpp1_coulomb}) are handled by the variable transformation \cite{Mohr:1974b}
\begin{equation}\label{eq_mpint0}
\left\{ \begin{array}{l}
y=2\gamma x_>\,,\\ 
r=x_</x_> \,, 
\end{array} \right.
\end{equation}
where $\gamma=\alpha Z$, $x_> = {\rm max}(x_1,x_2)$, $x_< = {\rm min}(x_1,x_2)$.

The integrals are done numerically using a $\tanh-\sinh$ quadrature scheme for the integrals over $r$ and $\exp-\sinh$ quadrature for the integral over $y$~\cite{Bailey:2005,Bailey:2024}. Concerning the integral over $z$, Gaussian quadratures were used for the high-energy part and $\tanh-\sinh$ ones for the low-energy part.

This is an important difference with respect to Ref.~\cite{Ferenc:2025}, where the radial integrals involving Gaussian basis functions were calculated analytically. Although it would be possible to do at least one radial integration analytically in our exponential basis set too, we have preferred to calculate all the integrals numerically. This choice is linked to the perspective of extending our method to one-electron molecular systems, where the number of radial integrations increases from two to four -- due to the loss of spherical symmetry -- so that it will not be possible to perform all of them analytically. It is therefore useful to study the impact of numerical integration, combined with that of using a numerical Green function, in order to assess the feasibility and computational requirements of high-precision calculations.

\subsection{Calculation of $\Delta E_{\kappa}^{1}$} \label{sec-g1}

The radial one-potential Green function $G_{\kappa}^1$ is calculated using \eq(\ref{eq_g1}) from the numerical free Green function $G_{\kappa}^0$, which is expressed similarly to \eq(\ref{gf1}):
\begin{equation}
[G_{\kappa}^0]_{i,j} (x_1,x_2,z) = \sum_{n=1}^{n_G} \frac{\phi_{\kappa,n}^{0,i}(x_1) \phi_{\kappa,n}^{0,j}(x_2)}{z - E_{\kappa,n}^0} \,,
\end{equation}
where $E_{\kappa,n}^0,\phi_{\kappa,n}^0$ are the eigenvalues and eigenvectors of the free radial Dirac equation, obtained with the same basis set as that used for the hydrogenic problem, and $n_G = 2N$ is the basis size. Injecting this expression into \eq(\ref{eq_g1}), we get, assuming $x_2<x_1$,
\begin{equation} \label{eq-integg11}
\begin{split}
&[G_{\kappa}^{1}]_{i,j}(r,y,z) = \int_0^{\infty} dx \, x^2 \sum_{k=1}^2 \sum_{m=1}^{n_G} \sum_{n=1}^{n_G} \\
&\frac{\phi_{\kappa,m}^{0,i}(y/(2\gamma))\phi_{\kappa,m}^{0,k}(x)}{z-E_{\kappa,m}^0} V(x)
\frac{\phi_{\kappa,n}^{0,k}(x)\phi_{\kappa,n}^{0,j}(ry/(2\gamma))}{z-E_{\kappa,n}^0}\,,
\end{split}
\end{equation}
where the change of variables of \eq(\ref{eq_mpint0}) was used.

The computational complexity of the calculation of $G_{\kappa}^1$ at a single point $(r,y,z)$ according to \eq(\ref{eq-integg11}) is $\mathcal{O}(n_G^3)$, where we use the ``big O" notation~\cite{Cormen:2022}. This is because the cost of calculating values of the $\phi_{\kappa,m}^{0,i}$ functions is $\mathcal{O}(n_G)$, and we recall that the integration is done analytically. The integrals over space and energy in \eq(\ref{eq_mpp1}) or (\ref{eq_mpp1_coulomb}) are instead evaluated numerically. Assuming that equal numbers of quadrature points $n_q$ are used for integration over the $r$, $y$ and $z$ variables, the calculation of $\Delta E_{\kappa}^{1}$ (see \eq(\ref{eq_Emany2p}) and explanation thereafter) has a computational time complexity of $\mathcal{O}(n_q^3 n_G^3)$. In this Section, we give an algorithm which reduces this computational complexity to at most $\mathcal{O}(\max(n_G, n_q)^4)$.

We start by defining
\begin{equation}\label{eq-integg13}
h_{\kappa,mn} = \sum_{k=1}^2\int_0^{\infty} dx \, x^2 \phi_{\kappa,m}^{0,k}(x)
V(x) \phi_{\kappa,n}^{0,k} (x)\,.
\end{equation}
\eq(\ref{eq-integg11}) can then be written as
\begin{equation}\label{eq-integg14}
[G_{\kappa}^{1}]_{i,j} (r,y,z) = \sum_{m,n=1}^{n_G}\frac{\phi_{\kappa,m}^{0,i}(y/(2\gamma)) h_{\kappa,mn} \phi_{\kappa,n}^{0,j}(ry/(2\gamma))}{(z-E_{\kappa,m}^0)(z-E_{\kappa,n}^0)}\,.
\end{equation}
With the definition
\begin{equation}\label{eq-integg14b}
v_{\kappa,m}^j(r,y,z) = \sum_{n=1}^{n_G}  \frac{h_{\kappa,mn}\phi_{\kappa,n}^{0,j}(ry/(2\gamma))}{(z-E_{\kappa,n}^0)(z-E_{\kappa,m}^0)}\,,
\end{equation}
\eq(\ref{eq-integg14}) becomes
\begin{equation}\label{eq-integg15}
[G_{\kappa}^{1}]_{i,j} (r,y,z) = \sum_{m=1}^{n_G} \phi_{\kappa,m}^{0,i}(y/(2\gamma)) v_{\kappa,m}^j(r,y,z)\,.
\end{equation}
The evaluation of $\phi_{\kappa,m}^{0,i}$ for a given $m$ and all quadrature points has a computational complexity of $\mathcal{O}(n_q^2 n_G)$. Since we need $\phi_{\kappa,m}^{0,i}$ for all values of $m$ for the evaluation of \eq(\ref{eq-integg15}), the associated computational complexity is $\mathcal{O}(n_q^2 n_G^2)$. The evaluation of $h_{\kappa,mn}$ for all values of $m$ and $n$ costs $\mathcal{O}(n_G^4)$, the integration over $x$ being done analytically. With the $h_{\kappa,mn}$ already evaluated and stored in arrays, the evaluation of $v_{\kappa,m}^j(r,z)$ costs $\mathcal{O}(n_q^2 n_G^2)$. This quantity is evaluated prior to integration and stored in an array. Finally, the integration in \eq(\ref{eq_mpp1}) or (\ref{eq_mpp1_coulomb}) using \eq(\ref{eq-integg15}) to evaluate $G_{\kappa}^1$ values costs $\mathcal{O}(n_q^3 n_G)$. Recapitulating, the successive steps of the evaluation have complexities of $\mathcal{O}(n_q^2 n_G^2)$, $\mathcal{O}(n_G^4)$, and $\mathcal{O}(n_q^3 n_G)$, hence the overall scaling $\mathcal{O}(\mathrm{max}(n_G,n_q)^4)$.

With this algorithm, the evaluation of $\Delta E_{\kappa}^{1}$ is only about 1.5 times longer than that of $\Delta E_{\kappa}$ or $\Delta E_{\kappa}^{0}$. This is a substantial improvement with respect to the computation times reported in~\cite{Ferenc:2025}, which are dominated by the evaluation of $\Delta E_{\kappa}^{1}$ (36 times longer than $\Delta E_{\kappa}$ or $\Delta E_{\kappa}^{0}$).

\section{Results} \label{sec-results}

We now present our results for the many-potential term of the SE correction. Firstly, we use a high-$Z$ atom (hydrogenlike uranium, $Z=92$) as a benchmark system in order to compare different basis sets. Secondly, we focus on the hydrogen atom ($Z=1$), which is the most relevant atomic system in the perspective of extending our method to molecular hydrogen ions, and present results obtained using the DKB exponential basis set introduced in Sec.~\ref{sec-dirac-eq}.

We give the results using the dimensionless function $F(Z\alpha)$ defined as
\begin{equation}
    \Delta E_{\rm SE} = \frac{\alpha}{\pi}(Z\alpha)^4 F(Z\alpha)\,.
\end{equation}

\subsection{Results for $Z=92$} \label{sec-z92}

In this section, all the calculations are performed in the Feynman gauge, i.e. using \eq (\ref{eq_mpp1}). Table~\ref{tab_km1} shows detailed results for each contribution of the many-potential term $\Delta E_{\kappa}^{2+}$ [see \eq (\ref{eq_Emany2p})] for $\kappa = \mp 1$, using different basis sets to solve the Dirac equation and calculate the Green function. The basis size is set to $n_G = 100$, except for the DKB2 basis set, where $n_G = 200$. A typical number of quadrature points is $268$, $304$ and $100$ for the $r$, $y$ and $z$ variables respectively. The evaluation of $\Delta E_{\kappa}^{2+}$ for a given value of $\kappa$ with $n_G=100$ and 16 cores takes roughly 20 minutes.

Our values are then compared to calculations performed using the analytical form of the Green function~\cite{Yerokhin:2025} (the detailed breakout of $\Delta E_{\kappa}^{2+}$ is given in~\cite{Ferenc:2025}, citing a private communication from the author of~\cite{Yerokhin:2025}). In view of the high precision of the latter results, the differences shown in the ``Diff.'' columns provide reliable estimates of the error.

Overall, our results clearly evidence a cancellation of errors, as observed in~\cite{Ferenc:2025}: the error of $\Delta E_{\kappa}^{2+}$ is much smaller than the individual errors of the zero-potential, one-potential and bound contributions. In order to quantify this effect, we give in Table~\ref{tab_km1} the values of the cancellation ratio $\rho$ defined as
\begin{equation} \label{eq_cancel}
\rho = \frac {\sqrt{\delta \left(\Delta E_{\kappa}\right)^2 \!+\! \delta\left(\Delta E_{\kappa}^0\right)^2 \!+\! \delta \left(\Delta E_{\kappa}^1\right)^2}} {\left| \delta \!\left(\Delta E_{\kappa}^{2+}\right) \right|},
\end{equation}
where $\delta (x)$ denotes the error on $x$.

The errors on the individual terms have the same order of magnitude (typically in the $10^{-3}$ range) for each of the tested basis sets, and no clear correlation is observed between these errors and the precision of the final result. This suggests that the cause for this low precision is not related to the specific characteristics of the basis set, but is common to all basis-set approaches. It is likely to be related to the fact that the zero-potential and bound Green functions have a discontinuity at the line $x_1 = x_2$~\cite{Yerokhin:2020} and are therefore poorly represented in the vicinity of this point in a finite basis, since the numerical Green function [\eq (\ref{gf1})] is continuous. On the other hand, the many-potential Green function has no discontinuity and does not suffer from this limitation.

In light of this, only the precision of the final result $\Delta E_{\kappa}^{2+}$ should be considered to evaluate the performance of a given basis set for SE calculations. Our results show that the RKB exponential basis set, for $\kappa = +1$, yields a substantial improvement (about one order of magnitude) with respect to the NKB one. We recall that for $\kappa = -1$ the NKB basis set actually respects the RKB condition. Overall, the DKB version is clearly superior, improving the precision further by two orders of magnitude. This is in agreement with the findings of~Ref.~\cite{Shabaev:2004}, where SE calculations in a $B$-spline basis set were reported for hydrogenlike calcium ($Z=20$).

One of the advantages of the DKB scheme is the elimination of spurious states~\cite{Shabaev:2004}. In practice, we observe spurious states in the Dirac spectrum for $\kappa>0$ when using the NKB basis set (but not when using RKB or DKB). For example, for $\kappa=1$, a $p_{1/2}$ state arises, whose energy coincides with that of the $1s_{1/2}$ state~\cite{Tupitsyn:2008}. We note that, although such states were not removed in our calculations of the numerical Green function, the precision gains that are obtained using DKB are similar for $\kappa=-1$, where no spurious states are expected nor observed, and for $\kappa = 1$. This indicates that the presence of spurious states is not the most important factor limiting the precision of the SE calculations.

The absolute precision for $n_G=100$ (``DKB'') is close to $10^{-6}$, and is slightly worse than that reported in~\cite{Ferenc:2025} using an RKB Gaussian basis set of the same size (``RKBG''). Increasing the basis size to $n_G = 200$ (``DKB2'') improves the precision by more than one order of magnitude. We postpone a full assessment of the respective merits of exponential and Gaussian basis sets to future work, since a thorough optimization of the basis set parameters has yet to be performed.

We extended our calculations to higher $|\kappa|$ values in order to determine the total value of the SE correction, see Table~\ref{tab_Z92}. The NKB and DKB2 basis sets were used for the low- and high-energy parts, respectively. The values of $\Delta E_{\kappa}^{2+}$ are compared to the highly precise values of Ref.~\cite{Yerokhin:2005}. Up to $|\kappa| = 15$, differences remain not larger than a few $10^{-7}$, but slowly increase with $\kappa$. The fact that the sum from $|\kappa| = 16$ to $35$ differs from the analytical Green function result by about $2 \times 10^{-5}$ indicates that this trend continues at higher $|\kappa|$. It can be attributed to the limitations of the basis set, since no significant loss of precision is observed in the numerical integration. The precision of our results could be improved by optimizing the nonlinear parameters of the basis set for each $\kappa$, which requires substantial effort.

Due to these growing uncertainties, it is preferable to use a limited number of partial waves in order to estimate the SE correction, and to extrapolate the remainder. The terms $|\kappa| = 16-20$ are fitted to polynomials in $1/|\kappa|$ as done in~\cite{Yerokhin:2025}, and the results of the least-squares fit are used to extrapolate the partial-wave sum to infinity. The uncertainty is estimated from the variation of the extrapolated sum as the maximum value of $|\kappa|$ is varied by 20\%, similarly to Ref.~\cite{Yerokhin:2025}. From Table~\ref{tab_Z92}, the total error induced by the finite basis set and by numerical integration on the sum of the first 20 terms is estimated to be smaller than $4\times 10^{-6}$ and is thus much smaller than the uncertainty from the extrapolation. The resulting uncertainty of the total SE correction is larger than that of the results obtained with the analytical Green function~\cite{Yerokhin:2005} by almost one order of magnitude, but slightly smaller than in the previous best results with a numerical Green function~\cite{Ferenc:2025}.

\begin{table*}[t]
\begin{tabular}{|c|c|l|r|l|l|l|r|l|c|c|}
\hline \hline
\multirow{2}{1.2em}{$\;\kappa$} & \multirow{2}{3.5em}{Method} & \multicolumn{2}{c|}{$\Delta E_{\kappa}$} & \multicolumn{2}{c|}{$\Delta E_{\kappa}^{0}$} & \multicolumn{2}{c|}{$\Delta E_{\kappa}^{1}$} & \multicolumn{2}{c|}{$\Delta E_{\kappa}^{2+}$} & \multirow{2}{3em}{$\;\;\;\;\rho$} \Tstrut \Bstrut \\
& & \multicolumn{1}{c}{Value} & \multicolumn{1}{c|}{Diff.} & \multicolumn{1}{c}{Value} & \multicolumn{1}{c|}{Diff.} & \multicolumn{1}{c}{Value} & \multicolumn{1}{c|}{Diff.} & \multicolumn{1}{c}{Value} & \multicolumn{1}{c|}{Diff.} & \\
\hline
\multirow{6}{2em}{$\,-1$}
& Anal.$\,$\cite{Yerokhin:2025priv} 
& $4.274\,776\,90$ & \multicolumn{1}{c|}{-} & $2.426\,503\,43$ &  \multicolumn{1}{c|}{-} & $\;\;\,0.336\,957\,11$ &  \multicolumn{1}{c|}{-} & $1.511\,316\,36$ &  \multicolumn{1}{c|}{-} &  \multicolumn{1}{c|}{-} \Tstrut \\
& RKB & $4.277\,423\,99$ & $2.6[-3]$ & $2.421\,135\,31$ & $-5.4[-3]$ & $\;\;\,0.345\,660\,87$ & $8.7[-3]$ & $1.510\,627\,81$ & $6.9[-4]$ & $1.5[+1]$ \\
& DKB  & $4.267\,152\,82$ & $-7.6[-3]$ & $2.416\,184\,74$ & $-1.0[-2]$ & $\;\;\,0.339\,650\,15$ & $2.7[-3]$ & $1.511\,317\,94$ & $1.6[-6]$ & $8.3[+3]$ \\
& DKB2 & $4.272\,991\,08$ & $-1.8[-3]$ & $2.424\,052\,12$ & $-2.5[-3]$ & $\;\;\,0.337\,622\,46$ & $6.7[-4]$ & $1.511\,316\,49$ & $1.3[-7]$ & $2.3[+4]$ \\
& RKBG$\,$\cite{Ferenc:2025}
& $4.271\,705\,9$ & $-3.1[-3]$ & $2.421\,600\,5$ & $-4.9[-3]$ & $\;\;\,0.338\,788\,8$ & $1.8[-3]$ & $1.511\,316\,6$ & $2.4[-7]$ & $2.5[+4]$\Bstrut \\
\hline
\multirow{6}{2em}{$\;\;1$}
& Anal.$\,$\cite{Yerokhin:2025priv}
& $2.201\,063\,89$ & \multicolumn{1}{c|}{-} & $2.584\,060\,05$ &  \multicolumn{1}{c|}{-} & $-0.503\,886\,64$ &  \multicolumn{1}{c|}{-} & $0.120\,890\,48$ &  \multicolumn{1}{c|}{-} &  \multicolumn{1}{c|}{-} \Tstrut \\
& NKB  & $2.186\,109\,46$ & $-1.5[-2]$ & $2.575\,627\,69$ & $-8.4[-3]$ & $-0.513\,550\,31$ & $-9.7[-3]$ & $0.124\,032\,09$ & $3.1[-3]$ & $6.3[+0]$ \\
& RKB  & $2.201\,796\,85$ & $\;\;\;7.3[-4]$ & $2.578\,401\,93$ & $-5.7[-3]$ & $-0.497\,752\,84$ & $6.1[-3]$ & $0.121\,147\,76$ & $2.6[-4]$ & $3.3[+1]$ \\
& DKB  & $2.193\,849\,36$ & $-7.2[-3]$ & $2.574\,254\,27$ & $-9.8[-3]$ & $-0.501\,297\,11$ & $2.6[-3]$ & $0.120\,892\,20$ & $1.7[-6]$ & $7.2[+3]$ \\
& DKB2 & $2.199\,298\,16$ & $-1.8[-3]$ & $2.581\,637\,23$ & $-2.4[-3]$ & $-0.503\,493\,36$ & $6.6[-4]$ & $0.120\,890\,54$ & $6[-8]$ & $5[+4]$ \\
& RKBG$\,$\cite{Ferenc:2025}
& $2.197\,906\,1$ & $-3.2[-3]$ & $2.579\,083\,4$ & $-5.0[-3]$ & $-0.502\,067\,9$ & $1.8[-3]$ & $0.120\,890\,7$ & $2.2[-7]$ & $2.8[+4]$\Bstrut \\
\hline \hline
\end{tabular}
\caption{Components of the many-potential term [\eq (\ref{eq_Emany2p})] for the ground state of hydrogenlike uranium ($Z=92$), using different basis sets to calculate the numerical Green function. For each value of $\kappa$, the first line is the reference value obtained with the analytical Green function, and in the following lines ``Diff.'' is the difference with respect to this value. The cancellation ratio $\rho$, defined in \eq (\ref{eq_cancel}), is given in the last column. All calculations are done with a basis size $n_G=100$, except for the ``DKB2'' basis set, where $n_G = 200$. $\alpha=1/137.035\,989\,5$.} \label{tab_km1}
\end{table*}

\begin{table*}[t]
\begin{tabular}{|l|r|r|r|}
\hline\hline
 \multicolumn{1}{|c|}{Term} & \multicolumn{1}{c|}{This work} & \multicolumn{1}{c|}{Anal.~\cite{Yerokhin:2005}} & \multicolumn{1}{c|}{Diff.} \Tstrut \Bstrut \\
\hline
$\Delta E_{\rm SE}^0\!+\!\Delta E_{\rm SE}^1$ & $-0.171\,545$\;\;\;\;\, & $-0.171\,545$\;\;\;\;\,
&  \Tstrut \Bstrut \\
\hline
\hspace{3mm}$|\kappa|=1$ & $1.632\,207\,03$\, & $1.632\,206\,83$\, & $2.0[-7]$\Tstrut \\
\hspace{11mm}$2$         & $0.012\,042\,45$\, & $0.012\,042\,27$\, & $1.8[-7]$ \\
\hspace{11mm}$3$         & $0.008\,313\,65$\, & $0.008\,313\,41$\, & $2.4[-7]$ \\
\hspace{11mm}$4$         & $0.003\,806\,13$\, & $0.003\,805\,81$\, & $3.2[-7]$ \\
\hspace{11mm}$5$         & $0.001\,988\,71$\, & $0.001\,988\,36$\, & $3.5[-7]$ \\
\hspace{11mm}$6$         & $0.001\,158\,32$\, & $0.001\,157\,88$\, & $4.4[-7]$ \\
\hspace{11mm}$7$         & $0.000\,730\,96$\, & $0.000\,730\,56$\, & $4.0[-7]$ \\
\hspace{11mm}$8$         & $0.000\,489\,98$\, & $0.000\,489\,62$\, & $3.6[-7]$ \\
\hspace{11mm}$9$         & $0.000\,344\,15$\, & $0.000\,343\,82$\, & $3.3[-7]$ \\
\hspace{10mm}$10$        & $0.000\,250\,83$\, & $0.000\,250\,55$\, & $2.8[-7]$ \\
\hspace{10mm}$11$        & $0.000\,188\,37$\, & $0.000\,188\,17$\, & $2.0[-7]$ \\
\hspace{10mm}$12$        & $0.000\,145\,00$\, & $0.000\,144\,88$\, & $1.2[-7]$ \\
\hspace{10mm}$13$        & $0.000\,113\,93$\, & $0.000\,113\,91$\, & $0.2[-7]$ \\
\hspace{10mm}$14$        & $0.000\,091\,09$\, & $0.000\,091\,17$\, & $-0.8[-7]$ \\
\hspace{10mm}$15$        & $0.000\,073\,91$\, & $0.000\,074\,10$\, & $-1.9[-7]$ \\
\hspace{8mm}$\sum_{|\kappa|=16}^{35}$ 
                         & $0.000\,399$\;\;\;\;\; & $0.000\,420$\;\;\;\;\; & $-2.1[-5]$\Bstrut \\
\hline
\hspace{8.5mm}$\sum_{|\kappa|=1}^{20}$   & $1.662\,16$\;\;\;\;\;\;\; &    & \Tstrut \Bstrut \\
$\sum_{|\kappa|=21}^{\infty}$ (extrap.)  & $0.000\,28(2)$\;\;        &    &  \\
\hspace{8mm}$\Delta E_{\rm SE}^{2+}$ & $1.662\,44(2)$\;\;    & $1.662\,461(3)$\; &  \Tstrut \Bstrut \\
\hline
\multicolumn{1}{|c|}{Total}              & $1.490\,90(2)$\;\; & $1.490\,916(3)$\; & \Tstrut \Bstrut \\\hline \hline
\end{tabular}
\caption{Individual contributions to the SE correction for the ground state of hydrogenlike uranium in the Feynman gauge. Results of calculations done with the analytical Green function are given in column 3, and the difference from these results is shown in the last column. $\alpha=1/137.035\,989\,5$.} \label{tab_Z92}
\end{table*}

\subsection{Results for $Z=1$}

We now turn to the hydrogen atom case, which is the most difficult and, to the best of our knowledge, has so far only been investigated using the analytical Green function~\cite{Jentschura:1999,Jentschura:2001,Yerokhin:2025}. In Ref.~\cite{Yerokhin:2025}, the refinements reviewed in Sec.~\ref{sec-se-eq}, i.e. application of convergence acceleration schemes and use of the Coulomb gauge, were shown to allow for substantial improvements in precision. In what follows, we study their impact on our basis-set calculations.

Table~\ref{tab_Z1_F} presents the results obtained using the Feynman gauge. Basis sizes ranging from $n_G=300$ to $600$ were used as the convergence is much slower than for $Z=92$. Comparison between columns 2 and 3 clearly evidences the acceleration of convergence of the partial-wave expansion permitted by the subtraction scheme proposed in~\cite{Sapirstein:2023}. The overall precision is limited by two main factors. Firstly, low-$\kappa$ contributions (especially $|\kappa|=1$, which is by far the largest) are difficult to calculate with high absolute precision, and their uncertainties can be estimated to about 10$^{-4}$. Secondly, although the absolute uncertainties decrease for higher values of $\kappa$ (in the 10$^{-5}$ range above $|\kappa| = 5$), the remainder after subtraction has a large relative uncertainty, requiring to start the extrapolation to $\kappa \to \infty$ at a low order ($|\kappa| = 21$) and making it unstable. This is why the extrapolation is the leading source of uncertainty. Using the results of~\cite{Yerokhin:2025} for 
the zero and one-potential terms (also for the Coulomb gauge results, see below), the total self-energy correction is eventually obtained with a relative uncertainty of $10^{-4}$.

The results obtained in the Coulomb gauge are shown in Table~\ref{tab_Z1_C}. The basis sizes used in these calculations range between $n_G = 600$ and $n_G = 640$. The absence of numerical cancellations - as already demonstrated in~\cite{Yerokhin:2025} - allows reaching higher accuracies for the individual $\kappa$ contributions, in the $10^{-5}-10^{-6}$ range. Again, the lowest-order contributions, $|\kappa|=1$ and $2$, are the least precise ones. The $|\kappa| = 3$ term is calculated with a much smaller error, after which the error increases with $|\kappa|$. It is dominated by the basis set limitation. As already mentioned for the $Z=92$ case (Sec.~\ref{sec-z92}), precision could be improved by optimizing the nonlinear parameters of the basis set for each $\kappa$. Here, we tested a different approach that consists in extrapolating the results to $n_G \to \infty$. For a range of $\kappa$ ($10 \leq |\kappa| \leq 15$), we performed calculations at $n_G =240$, $360$, $480$, and $600$, and fitted the results by a power law ($a+b/n_G^c$). This allows for a clear reduction of the errors, and the remaining discrepancies with the result obtained with the analytical Green function are compatible with the uncertainties arising from the fitting procedure.

The improved absolute accuracies and accelerated convergence with respect to the Feynman gauge allow for a more precise determination of the total SE correction. However, these advantages also imply the need for even higher accuracies (for the same $|\kappa|$) to make the $|\kappa| \to \infty$ extrapolation reliable. Due to this, we have to start using the extrapolation at a lower value ($|\kappa|=11$) as compared to the Feynman gauge, which strongly limits the accuracy. Despite this limitation, the precision of the total self-energy correction is improved by one order of magnitude, reaching $10^{-5}$.

\begin{table}[]
\begin{tabular}{|l|r|r|}
\hline\hline
\multicolumn{1}{|c|}{$\Delta E_{\rm SE}^0$~(Ref.\,\cite{Yerokhin:2025})} & \multicolumn{2}{c|}{$-168\,176.156\,251$} \Tstrut \\
\multicolumn{1}{|c|}{$\Delta E_{\rm SE}^1$~(Ref.\,\cite{Yerokhin:2025})}  & \multicolumn{2}{c|}{$\phantom{-}148\,579.466\,946$} \Bstrut \\
\hline
& Without subtr. & With subtr. \Tstrut \Bstrut \\
\hline
\multicolumn{1}{|c|}{$\Delta E_{\rm substr}^{2+}$} &  & $9\,370.581\,414$\Tstrut \\
\hspace{3mm}$|\kappa|=1$ & $19\,507.536\,257$ & $10\,234.905\,709$\\
\hspace{11mm}$2$ & $58.961\,982$ & $1.345\,695$ \\
\hspace{11mm}$3$ & $15.122\,261$ & $0.109\,036$ \\
\hspace{11mm}$4$ & $7.177\,115$  & $0.031\,781$ \\
\hspace{11mm}$5$ & $4.215\,621$  & $0.013\,579$ \\
\hspace{11mm}$6$ & $2.768\,017$  & $0.007\,010$ \\
\hspace{11mm}$7$ & $1.951\,357$  & $0.004\,060$ \\
\hspace{11mm}$8$ & $1.443\,749$  & $0.002\,542$ \\
\hspace{11mm}$9$ & $1.107\,123$  & $0.001\,652$ \\
\hspace{11mm}$10$ & $0.872\,815$ & $0.001\,125$ \\
\hspace{11mm}$11$ & $0.703\,439$ & $0.000\,794$ \\
\hspace{11mm}$12$ & $0.577\,242$ & $0.000\,576$ \\
\hspace{11mm}$13$ & $0.480\,851$ & $0.000\,427$ \\
\hspace{11mm}$14$ & $0.405\,684$ & $0.000\,320$ \\
\hspace{11mm}$15$ & $0.346\,028$ & $0.000\,243$\Bstrut \\
\hline
\hspace{8mm}$\sum_{|\kappa|=16}^{20}$  &  & $0.000\,572$\Tstrut \\
\multicolumn{1}{|c|}{$\sum_{|\kappa|=21}^{\infty}$ (extrap.)} &  & $0.000(1)\;$ \\
\multicolumn{1}{|c|}{$\Delta E_{\rm SE}^{2+}$}     &  &$19\,607.007(1)\;$\Bstrut \\
\hline
\multicolumn{1}{|c|}{Total} &  & $10.318(1)\;$\Tstrut\Bstrut \\
\hline\hline
\end{tabular}
\caption{Individual contributions to the SE correction for the ground state of the hydrogen atom ($Z=1$) in the Feynman gauge. Columns 2 and 3 are calculated with and without applying the subtraction scheme described in Sec.~\ref{sec_accel}, respectively. $\alpha=1/137.036$. The final result in the last line should be compared to the reference result obtained with the analytical Green function~\cite{Jentschura:1999,Jentschura:2001}: $F(Z\alpha)=10.316\,793\,650(1)$.} \label{tab_Z1_F}
\end{table}

\begin{table}[]
\begin{tabular}{|l|r|r|r|}
\hline\hline
\multicolumn{1}{|c|}{Term} & \multicolumn{1}{c|}{This work} & \multicolumn{1}{c|}{Anal.\cite{Yerokhin:2025}} & \multicolumn{1}{c|}{Diff.}\Tstrut\Bstrut \\
\hline
\multicolumn{1}{|c|}{$\Delta E_{\rm SE}^0$} &  & $13.849\,474\,1$ & \Tstrut\\
\multicolumn{1}{|c|}{$\Delta E_{\rm SE}^1$}  &  & $-2.879\,681\,6$ & \Bstrut \\
\hline
\multicolumn{1}{|c|}{$\Delta E_{\rm substr}^{2+}$} &    
$-1.127\,787\,5$  & $-1.127\,787\,5$ & $0\;\;\;\;\;\;$\Tstrut\Bstrut\\
\hline
\hspace{3mm}$|\kappa|=1$ &   
$0.475\,589\,3$ &  $0.475\,625\,7$ & $-3.6[-5]$\Tstrut \\
\hspace{11mm}$2$          & 
$-0.004\,692\,5$ & $-0.004\,723\,6$ & $3.1[-5]$ \\
\hspace{11mm}$3$          &  
$0.001\,965\,1$ &  $0.001\,962\,3$ & $2.8[-6]$ \\
\hspace{11mm}$4$          &
$0.000\,790\,1$ &  $0.000\,785\,7$ & $4.4[-6]$ \\
\hspace{11mm}$5$          & 
$0.000\,401\,4$ & $0.000\,396\,6$ & $4.8[-6]$ \\
\hspace{11mm}$6$          &
$0.000\,233\,5$ & $0.000\,228\,1$ & $5.4[-6]$ \\
\hspace{11mm}$7$          & 
$0.000\,148\,9$ & $0.000\,142\,8$ & $6.1[-6]$ \\
\hspace{11mm}$8$          & 
$0.000\,101\,8$ & $0.000\,094\,8$ & $7.0[-6]$ \\
\hspace{11mm}$9$          &
$0.000\,073\,3$ & $0.000\,065\,8$ & $7.5[-6]$\Bstrut \\
\hline
\multirow{2}{1em}{$\hspace{11mm}10$} &  
$0.000\,055\,7$ & \multirow{2}{4.6em}{$0.000\,047\,2$} & $8.5[-6]$\Tstrut \\
& $0.000\,042\,6$ & & $-4.6[-6]$\Bstrut \\
\hline
\multirow{2}{1em}{$\hspace{11mm}11$} &
$0.000\,043\,8$ & \multirow{2}{4.6em}{$0.000\,034\,8$} & $9.0[-6]$\Tstrut \\
& $0.000\,031\,0$ & & $-3.8[-6]$\Bstrut \\
\hline
\multirow{2}{1em}{$\hspace{11mm}12$} &
$0.000\,036\,1$ & \multirow{2}{4.6em}{$0.000\,026\,3$} & $9.8[-6]$\Tstrut \\
& $0.000\,021\,1$ & & $-5.2[-6]$\Bstrut \\
\hline
\multirow{2}{1em}{$\hspace{11mm}13$} &
$0.000\,030\,7$ & \multirow{2}{4.6em}{$0.000\,020\,2$} & $1.0[-5]$\Tstrut \\
& $0.000\,016\,3$ &  & $-3.9[-6]$\Bstrut \\
\hline
\multirow{2}{1em}{$\hspace{11mm}14$} &
$0.000\,027\,0$ & \multirow{2}{4.6em}{$0.000\,015\,8$} & $1.1[-5]$\Tstrut \\
& $0.000\,009\,5$ & & $-6.3[-6]$\Bstrut \\
\hline
\multirow{2}{1em}{$\hspace{11mm}15$} &
$0.000\,024\,4$ & \multirow{2}{4.6em}{$0.000\,012\,5$} & $1.2[-5]$\Tstrut \\
& $0.000\,007\,9$ & & $-4.6[-6]$\Bstrut \\
\hline
\multicolumn{1}{|c|}{$\sum_{|\kappa|=11}^{\infty}$ (extrap.)} &  $0.000\,2(1)\;$  & & \Tstrut \\
\multicolumn{1}{|c|}{$\Delta E_{\rm SE}^{2+}$}                 &  $-0.652\,9(1)\;$ & & \Bstrut \\
\hline
\multicolumn{1}{|c|}{Total}                                 &   $10.316\,9(1)\;$ & $10.316\,794(1)$& \Tstrut\Bstrut \\
\hline \hline
\end{tabular}
\caption{Individual contributions to the many-potential term of the SE correction for the ground state of the hydrogen atom in the Coulomb gauge, using the convergence acceleration scheme described in Sec.~\ref{sec_accel}. Results of calculations done with the analytical Green function are given in column 3, and the difference from these results is shown in the last column. For $10 \leq |\kappa| \leq 15$, the first line is the value obtained with the largest basis size, whereas the second line is the value resulting from extrapolation to infinite basis size (see text for details). $\alpha=1/137.036$.} \label{tab_Z1_C}
\end{table}

\section{Conclusion}

We have presented numerical calculations of the SE correction in hydrogenlike atoms using a numerical Green function calculated in a finite basis set, and performing all the integrals numerically. Our results at high $Z$ indicate that the performance of the DKB exponential basis set used in our calculations is comparable to that of the Gaussian basis set used in~\cite{Ferenc:2025}. Further, the implementation of a convergence acceleration scheme allowed us to extend our calculations to the most difficult case $Z=1$, where a relative precision of $10^{-4}$ is achieved for the total SE correction in the Feynman gauge. An even higher precision in the calculation of partial-wave contributions to the many-potential term is achieved in the Coulomb gauge, allowing to improve the accuracy of the SE correction by one order of magnitude. The main limitation comes from the extrapolation to an infinite number of partial waves, and further improvements in accuracy are required to take full advantage of the known advantages of the Coulomb gauge for low-$Z$ calculations. A logical next step would consist in a dedicated optimization of the basis set parameters. The best way to perform this optimization is at present an open question, since the quantities relevant to the SE calculation do not obey a minimization principle. Nevertheless, these results represent a substantial step towards SE calculations in non-hydrogenic systems.

\vspace{2mm}

\textbf{Acknowledgements.} We thank V. A. Yerokhin, D. Ferenc and T. Saue for helpful discussions, and M. Panet for reading of the manuscript. This work was part of 23FUN04 COMOMET that has received funding from the European Partnership on Metrology, co-financed by the European Union’s Horizon Europe Research and Innovation Program and by the Participating States. Funder ID: 10.13039/100019599.

\appendix

\section{Dirac equation with different kinetic balances} \label{app_Dirac}

In this Appendix, we give the matrix form of the Dirac equation for the NKB, RKB and DKB schemes (see also~\cite{Sun:2011,Salman:2020}).

\subsection{NKB}

Injecting the expressions~(\ref{eq-basis-exp}) in \eq(\ref{eq_Dirac_rad}) leads to the following matrix representation of the Dirac equation:
\begin{equation}
\begin{pmatrix}
\mathrm{V}^{LL} & \mathrm{\Pi}^{LS} \\
\mathrm{\Pi}^{SL} & \mathrm{V}^{SS}\!-\! 2c^2 \mathrm{S}^{SS}
\end{pmatrix}
\begin{pmatrix}
A \\ B    
\end{pmatrix}
=
\begin{pmatrix}
\mathrm{S}^{LL} & 0 \\
0 & \mathrm{S}^{SS}
\end{pmatrix}
\begin{pmatrix}
A \\ B    
\end{pmatrix},
\end{equation}
with, assuming $X=L$ or $S$,
\begin{subequations}
\begin{equation}
[\mathrm{S}^{XX}]_{\mu\nu} = \int_0^{\infty} \pi_{\mu}^X(x) \pi_{\nu}^X(x) dx,
\end{equation}
\begin{equation}
[\mathrm{V}^{XX}]_{\mu\nu} = \int_0^{\infty} \pi_{\mu}^X(x) V(x) \pi_{\nu}^X(x) dx,
\end{equation}
\begin{equation}
[\mathrm{\Pi}^{LS}]_{\mu\nu} = \int_0^{\infty} \pi_{\mu}^L(x) \left[ -\frac{d}{dx} + \frac{\kappa}{x} \right] \pi_{\nu}^S(x) dx,
\end{equation}
\begin{equation}
[\mathrm{\Pi}^{SL}]_{\mu\nu} = \int_0^{\infty} \pi_{\mu}^S(x) \left[ \frac{d}{dx} + \frac{\kappa}{x} \right] \pi_{\nu}^L(x) dx.
\end{equation}
\end{subequations}

\subsection{RKB}

The RKB relationship~(\ref{eq_basis_rkb}) leads to the matrix equation:
\begin{equation}
\begin{pmatrix}
\mathrm{V}^{LL} & \mathrm{T}^{LL} \\
\mathrm{T}^{LL} & \frac{1}{4c^2} \mathrm{W}^{LL}\!-\! \mathrm{T}^{LL}
\end{pmatrix}
\begin{pmatrix}
A \\ B    
\end{pmatrix}
=
\begin{pmatrix}
\mathrm{S}^{LL} & 0 \\
0 & \frac{1}{2c^2} \mathrm{T}^{LL}
\end{pmatrix}
\begin{pmatrix}
A \\ B    
\end{pmatrix},
\end{equation}
where the matrix elements are given by
\begin{subequations}
\begin{equation}
[\mathrm{T}^{LL}]_{\mu\nu} = -\frac{1}{2} \int_0^{\infty} \pi_{\mu}^L(x) \left[ \frac{d^2}{dx^2} - \frac{\kappa(\kappa+1)}{x^2} \right] \pi_{\nu}^L(x) dx,
\end{equation}
\begin{equation}
[\mathrm{W}^{LL}]_{\mu\nu} = \int_0^{\infty} \left[ \frac{d\pi_{\mu}^L}{dx} + \frac{\kappa}{x} \pi_{\mu}^L \right] V(x) \left[ \frac{d\pi_{\nu}^L}{dx} + \frac{\kappa}{x} \pi_{\nu}^L \right] dx.
\end{equation}
\end{subequations}

\subsection{DKB}

With the DKB expansion~(\ref{eq_basis_dkb}), the matrix form of the Dirac equation is
\begin{equation}
\begin{split}
&\begin{pmatrix}
\mathrm{T}^{LL}\!+\!\mathrm{V}^{LL}\!+\!\frac{1}{4c^2} \mathrm{W}^{LL} & \frac{1}{2c} \left( \mathrm{W}^{LS}_a + \mathrm{W}^{LS}_b \right) \\
\frac{1}{2c} \left( \mathrm{W}^{SL}_a + \mathrm{W}^{SL}_b \right) & \mathrm{V}^{SS}\!-\!2 \mathrm{T}^{SS}\!+\! \frac{1}{4c^2} \mathrm{W}^{SS}\!-\!2c^2 \mathrm{S}^{SS}
\end{pmatrix} \\
& \;\; \times
\begin{pmatrix}
A \\ B    
\end{pmatrix}
=
\begin{pmatrix}
\mathrm{S}^{LL}\!+\!\frac{1}{2c^2} \mathrm{T}^{LL} & 0 \\
0 & \mathrm{S}^{SS}\!+\!\frac{1}{2c^2} \mathrm{T}^{SS}
\end{pmatrix}
\begin{pmatrix}
A \\ B    
\end{pmatrix},
\end{split}
\end{equation}
with
\begin{subequations}
\begin{equation}
[\mathrm{T}^{SS}]_{\mu\nu} = -\frac{1}{2} \int_0^{\infty} \pi_{\mu}^S(x) \left[ \frac{d^2}{dx^2} - \frac{\kappa(\kappa-1)}{x^2} \right] \pi_{\nu}^S(x) dx,
\end{equation}
\begin{equation}
[\mathrm{W}^{SS}]_{\mu\nu} = \int_0^{\infty} \left[ \frac{d\pi_{\mu}^S}{dx} - \frac{\kappa}{x} \pi_{\mu}^S \right] V(x) \left[ \frac{d\pi_{\nu}^S}{dx} - \frac{\kappa}{x} \pi_{\nu}^S \right] dx,
\end{equation}
\begin{equation}
\begin{split}
[\mathrm{W}^{LS}_a]_{\mu\nu} =& \int_0^{\infty} \pi_{\mu}^L V(x) \left[ \frac{d\pi_{\nu}^S}{dx} \!-\! \frac{\kappa}{x} \pi_{\nu}^S \right] dx \\
&+\int_0^{\infty} \left[ \frac{d\pi_{\mu}^L}{dx} \!+\! \frac{\kappa}{x} \pi_{\mu}^L \right] V(x) \pi_{\nu}^S dx,
\end{split}
\end{equation}
\begin{equation}
[\mathrm{W}^{LS}_b]_{\mu\nu} = \frac{1}{2} \int_0^{\infty} \left[ \frac{d\pi_{\mu}^L}{dx} \!+\! \frac{\kappa}{x} \pi_{\mu}^L \right] \left[ \frac{d^2\pi_{\nu}^S}{dx^2} \!-\! \frac{\kappa(\kappa-1)}{x^2} \pi_{\nu}^S \right]  dx,
\end{equation}
\begin{equation}
\begin{split}
[\mathrm{W}^{SL}_a]_{\mu\nu} =& \int_0^{\infty} \pi_{\mu}^S V(x) \left[ \frac{d\pi_{\nu}^L}{dx} \!+\! \frac{\kappa}{x} \pi_{\nu}^L \right] dx \\
&+\int_0^{\infty} \left[ \frac{d\pi_{\mu}^S}{dx} \!-\! \frac{\kappa}{x} \pi_{\mu}^S \right] V(x) \pi_{\nu}^L dx,
\end{split}
\end{equation}
\begin{equation}
[\mathrm{W}^{SL}_b]_{\mu\nu} = \frac{1}{2} \int_0^{\infty} \left[ \frac{d\pi_{\mu}^L}{dx} \!-\! \frac{\kappa}{x} \pi_{\mu}^S \right] \left[ -\frac{d^2\pi_{\nu}^L}{dx^2} \!+\! \frac{\kappa(\kappa+1)}{x^2} \pi_{\nu}^L \right]  dx.
\end{equation}
\end{subequations}

\end{document}